%% file: nep-thm.tex
\documentclass[epj, preprint]{svjour}
\usepackage{graphics}
\usepackage{natbib}
\usepackage{xspace}
\usepackage{amssymb,amsmath}

\def \e { \mbox{$\mathrm{e}$} }

\newcommand\eg{\textit{e.g.}\xspace}
\newcommand\ie{\textit{i.e.}\xspace}
\newcommand{\alf}{Alfv$\acute{\text{e}}$n\xspace}
\renewcommand\Re{\mbox{$\mathrm{Re}$}}
\renewcommand\Im{\mbox{$\mathrm{Im}$}}

\newcommand{\vkt}[1]{\tilde{#1}}

\newcommand\vkf{\vkt{f}}
\newcommand \vkS{\vkt{S}}
\newcommand \vks{\vkt{s}}
\renewcommand \S{S}
\renewcommand \L{L}
\def \vth {\mbox{$v_{\mathrm{T}}$}}
\def \omegaNL {\mbox{$\omega_{\mathrm{NL}}$}}

\begin{document}
\title{Irreversible energy flow in forced Vlasov dynamics}
\author{G. G. Plunk\inst{1}\thanks{gplunk@ipp.mpg.de}\and J. T. Parker\inst{2}}
%
\institute{Max Planck Institute for Plasma Physics, EURATOM Association \and Oxford Centre for Collaborative Applied Mathematics, University of Oxford}
%
%
\abstract{
The recent paper of \citet{plunk-landau} considered the forced linear Vlasov equation as a model for the quasi-steady state of a single stable plasma wavenumber interacting with a bath of turbulent fluctuations.  This approach gives some insight into possible energy flows without solving for nonlinear dynamics.  The central result of the present work is that the forced linear Vlasov equation exhibits asymptotically zero (irreversible) dissipation to all orders under a detuning of the forcing frequency and the characteristic frequency associated with particle streaming.  We first prove this by direct calculation, tracking energy flow in terms of certain exact conservation laws of the linear (collisionless) Vlasov equation.  Then we analyze the steady-state solutions in detail using a weakly collisional Hermite-moment formulation, and compare with numerical solution.  This leads to a detailed description of the Hermite energy spectrum, and a proof of no dissipation at all orders, complementing the collisionless Vlasov result.
\PACS{
      {PACS-key}{discribing text of that key}   \and
      {PACS-key}{discribing text of that key}
     } 
} 
\maketitle
\section{Introduction}
\label{intro-sec}

In this paper we revisit the equation considered by \citet{plunk-landau}, \ie the forced one-dimensional Vlasov equation for an electrostatic quasi-neutral plasma.  In studying this equation, we hope to gain insight into the more difficult problem of how Landau damping behaves in a turbulent setting.  The basic premise is as follows.  If a single linearly stable plasma Fourier component is participating in a turbulent steady state, involving many other wavenumbers, then the nonlinear term in the equation for this component must behave as a statistically stationary source, \ie it is a signal whose statistics do not depend on time.  

We thus consider the nonlinear term as a given random source.  The goal is not to describe the turbulent state in full, but rather to gain some insight into how Landau damping might interact with turbulence.  We stress that our calculation only loses generality when additional assumptions (beyond stationarity) are made about the properties of the source term.  In particular, nothing is assumed that would confine the applicability to weakly nonlinear regimes (``weak turbulence'').  Finally, note that by the Wiener--Khinchin theorem, the Fourier components (in time) of a stationary signal are uncorrelated, and therefore the statistical properties of the plasma component can be deduced from its exact response to a single frequency drive -- so-called ``harmonic forcing''.  Thus, the analysis of this paper will largely be carried out in the frequency domain.

This work originates from an interest in gyrokinetic turbulence, in which the free streaming of plasma particles along a magnetic guide field gives rise to linear Landau damping.  However, this occurs simultaneously to the nonlinear cascade of ``free energy'' that interferes with the Landau damping.  The gyrokinetic equation is applied to fusion experiments, but also to some astrophysical systems as well.  However, the equations in this paper are simpler than the full gyrokinetic system, and so it is possible that the results have a more general applicability.


To solve our problem, we will apply two different limits.  These are the collisionless and weakly collisional limits.  It is useful to discuss why the results are compatible.  In both cases Landau damping can be interpreted as an irreversible process.  In the collisionless picture, the linear process is irreversible in the sense that the damping occurs indefinitely with no recurrences.  That is, the density is never revived, despite the time-reversal symmetry of the Vlasov equation.  This irreversible behavior does not require explicit dissipation, and is made possible in the collisionless case by the fact that the Vlasov equation is infinite-dimensional (because the particle velocity is a continuous variable).  This enables the creation of arbitrarily fine scales in velocity space during the decay process.  In the weakly collisional case, a second (perhaps more physical) interpretation can be drawn, owing to the fact that, in a weakly collisional plasma, Landau damping is inextricably linked with collisional dissipation.  That is, the Boltzmann equation (Vlasov equation plus collisions) satisfies Boltzmann's H-theorem, whereby a specific (information theoretic) definition of entropy is a strictly increasing function of time.  Physically, the ``phase-mixing'' process involved in Landau damping induces a flow of energy to small scales where in any finite representation (\ie a numerical implementation) it must be ``mopped up'' by explicit dissipation like a collision operator.  Thus, there is a clear correspondence between the flux of energy (in velocity-scale-space) and the collisional entropy production of Boltzmann's H-theorem.

\begin{figure}
\resizebox{0.5\textwidth}{!}{%
  \includegraphics{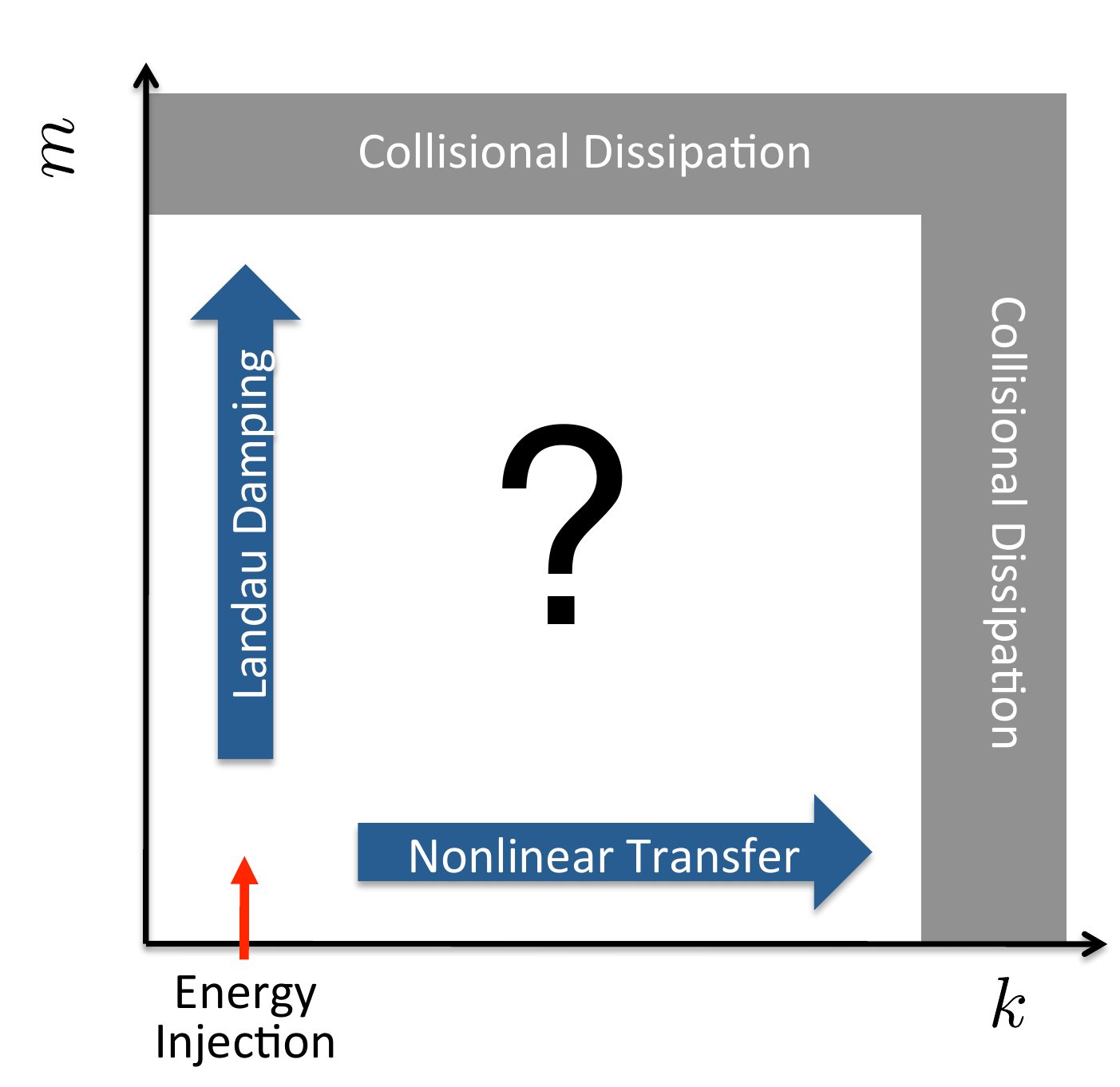}
}
\caption{Cartoon of turbulent energy flow via linear and nonlinear channels.}\label{cartoon-fig}
\end{figure}

In a turbulent setting, the problem of Landau damping can be stated as a problem of dissipation.  The basic question is: where does the energy go?  Energy is injected by some source, is nonlinearly redistributed among wavenumbers, exciting waves (sometimes these are identifiable as linear modes, but they need not be in strongly nonlinear regimes) and other degrees of freedom, and is ultimately routed to some dissipation channels.  Energy can be nonlinearly transferred between spatial scales (\ie cascaded in the sense of fluid turbulence), but it can also be redistributed within a single Fourier mode, \ie it can be shuffled between scales in velocity space.  For present purposes, the latter occurs by linear coupling.  Thus, we may conceptually think of the energy flow as occurring in a two dimensional space, as in Fig.~\ref{cartoon-fig}, where the horizontal axis represents the dimensions through which nonlinear flow of energy occurs, \eg the wavenumber $k$, and the vertical axis represents an index $m$ for the modes excited by the Landau damping process (\eg the Hermite index).  Note that this problem has been investigated recently \citep{hatch-2013, hatch-2014}, by direct numerical simulation of turbulence, in the context of fusion plasmas.\footnote{For their problem, \citet{hatch-2013, hatch-2014} found that the relative amount of energy dissipated at high $m$ is small at weak collisionality, in contrast with the behavior of linear Landau damping.  This establishes an intriguing connection with our work, which could be explored in the future.}

On basic theoretical grounds, it is unknown how the energy will in general flow in the plane of Fig.~\ref{cartoon-fig}, but one can imagine two extremes.  One was described by \citet{howes2008} (but, it should be noted, merely for the purpose of illustrating disagreement with observations).  It can be summarized as follows.  Since the flow of energy in this space is conservative, it can be represented by a simple continuity equation if it occurs locally (among neighboring wavenumbers).  The direct integration of the equation, under the assumption that linear damping always competes with the nonlinear transfer rate, then leads to an exponentially decaying spectrum of energy in $k$.  In the limit of weak collisionality, all of the energy flux is consumed by Landau damping, and no energy reaches large-$k$ dissipation mechanisms.  However, this exponential decay is not observed and instead a power-law spectrum persists.

On the other hand, one could reason that no energy should flow into the Landau channel.  This was argued in several papers including \citet{schek-ppcf, plunk-jfm}, based on the fact that nonlinear energy transfer accelerates with $k$ and so reaches high-$k$ dissipation more quickly than high-$m$ dissipation.  The weakness in this argument is that damping is determined by energy flux, not the speed of flow in $m$-space.  The fact that energy travels more slowly in the Landau channel can be attributed to shallowness of the spectrum.  Indeed, when the spectrum of energy is non-integrable (\eg $W(m) \sim m^{-n}$, $n \leq 1$) then energy can never dissipate in the limit of small collisionality; but {\it damping} would persist because the flux of energy through a finite-$m$ mode does not strongly depend on where or how the dissipation is achieved.

Clearly, simple arguments are not satisfactory to explain the role of Landau damping in kinetic plasma turbulence, and the issue remains fundamentally unresolved.  The only thing that can be concluded {\it a priori} is that if the spectrum of turbulence is modified by energy loss to the Landau channel, then the resulting turbulent spectrum (\ie the energy density in $k$-space) must be steeper than the unmodified spectrum.  

In short, the discussion above motives us to look for simple mechanisms that can obstruct or enhance the Landau damping process.  Linear Landau damping describes the decay of a single isolated wave that is left undisturbed for a sufficiently long time to approach a state of steady decay, whereby energy is drained from the wave via a route through successively smaller scales in velocity space.  How is this process affected when disrupted by the presence of an external source?  The remainder of the paper addresses this question.

\section{Equations and basic analysis}\label{eqns-sec}

We consider the one-dimensional electrostatic reduced Vlasov equation for a single Fourier component, under harmonic forcing

\begin{equation}
\frac{\partial f}{\partial t} + i k v f + i k c_s^2 n(t) G(v) = \exp(-i \Omega t) S(v),\label{plasma-eqn}
\end{equation}

\noindent where $k > 0$ for simplicity.  The density of $f$ is

\begin{equation}
n(t) = \int_{-\infty}^{\infty} dv f (v, t).\label{n-eqn}
\end{equation}

\noindent This system can be derived from the full kinetic system (\eg gyrokinetics) by integrating the kinetic equation over velocity coordinates that are not associated with the Landau damping.  Note that for simplicity, we are neglecting magnetic fluctuations.  However, in Appx.~\ref{KAW-sec} we demonstrate how to generalize the analysis to include a fluctuating magnetic field.  

We now apply the Morrison transform \citep{morrison-pfirsch, morrison-aa} to Eqn.~\ref{plasma-eqn} using the notation of \citet{plunk-landau}:

\begin{equation}
\frac{\partial \vkf}{\partial t} + i ku \vkf = \exp(-i \Omega t) \vkS(u),\label{plasma-eqn-trans}
\end{equation}

\noindent where the transformed distribution is defined

\begin{equation}
\vkf(u) = \frac{f_{+}(u)}{D_{+}(u)} + \frac{f_{-}(u)}{D_{-}(u)},
\end{equation}

\noindent and we define $D_{\pm}(u) = 1 \pm 2\pi i c_s^2 G_{\pm}(u)$.  The positive- and negative-frequency parts of an arbitrary function $h(v)$ are defined in terms of the Fourier transform by

\begin{equation}
h_{\pm}(u) = \pm \int_0^{\pm \infty} d\nu \e^{i \nu u} \int_{-\infty}^{\infty} dv\frac{\e^{-i \nu v}}{2 \pi} h(v).
\end{equation}

\noindent Note that $h(u) = h_{+}(u) + h_{-}(u)$ and $D_{+} = D_{-}^*$ with superscript $*$ denoting the complex conjugate.  We will also need the Hilbert transform, which for an arbitrary function $h$ is denoted with a subscript $*$:

\begin{equation}
h_*(u) = \frac{1}{\pi}P \int_{-\infty}^{\infty} \frac{h(v) dv}{u - v}.\label{hilbert-def}
\end{equation}

\noindent where $P$ denotes the principal value.  Following \citet{plunk-landau} the long-time ($t \rightarrow \infty$) solution of Eqn.~\ref{plasma-eqn-trans} can be written as

\begin{equation}
\vkf(u, t) = \exp(- i \Omega t)  \vkS(u) K(u - \Omega/k)/k,\label{fvk-large-t-eqn}
\end{equation}

\noindent where 

\begin{equation}
K(x) = \pi\left[\delta(x) - (i/\pi)P\frac{1}{x}\right],\label{K-eqn}
\end{equation}

\noindent which, using identity \ref{pm-hilbert-identity}, implies that $\int_{-\infty}^{\infty} dx K(x - y) h(x) = \pi (h + i h_*) = 2\pi h_+(y)$ for arbitrary $h$.  Then we may compute $f(v)$ by the inverse of the Morrison transform:

\begin{equation}
f(v, t) = \int_{-\infty}^{\infty} du \vkf(u, t) f^{u}(v),\label{vkt-inverse}
\end{equation}

\noindent where $f^{u}(v) = \lambda_u\delta(u - v) + c_s^2 P [G(v)/(u - v)]$, where $P$ denotes the principal value with respect to the point $u = v$, and $\lambda_u = 1 - c_s^2 \pi G_*(u)$.  We thus find

\begin{multline}
f(v, t) = \exp(- i \Omega t)\{(1-\pi c_s^2 G_*)\vkf_{\infty} - \pi c_s^2 G \vkf_{\infty*}\},\label{f-large-t-eqn}
\end{multline}

\noindent where $\vkf_{\infty} = \vkS(v) K(v - \Omega/k)/k$.

\section{``No Entropy Production'' (NEP) Theorem}
\label{nep-sec}

In \citet{plunk-landau}, damping was quantified by the wave-particle interaction term (\ie the electric field multiplied by the plasma current) directly.  Although this is an intuitive measure, and allows a direct comparison with Landau damping, it is illuminating to consider energy flow in a more general sense, via the conservative properties of the Vlasov system.  In the absence of forcing it is easy to see that 

\begin{equation}
\mathcal{W}(u) = |\vkf|^2/2,\label{Wcal-def}
\end{equation}

\noindent is conserved, as is any quantity formed from $\mathcal{W}$ by arbitrarily weighted integration over $u$.  In the presence of a source, the balance equation is

\begin{equation}
\begin{split}
\frac{d\mathcal{W}}{dt}& = \Re[\vkf^* \vkS]\\
& = \frac{1}{k}|\vkS|^2 \Re[K(u - \Omega/k)]\\
& = \frac{\pi}{k}|\vkS|^2 \delta(u - \Omega/k),
\end{split}\label{Wcal-balance-eqn}
\end{equation}

\noindent where we have used the steady state solution for $\vkf$, Eqn.~\ref{fvk-large-t-eqn}, between the first line and second line, and $\Re[K(x)] = \delta(x)$ between the second line and third line.  Unsurprisingly, this quasi-steady state is singular, but by (arbitrarily) integrating over $u$ we can extract well-defined measures of input ``power''.  Such integrals generally fall off sharply at small $k\vth/\Omega$ as expected from the suppression of damping found in \citet{plunk-landau}.  This can be seen by rewriting

\begin{equation}\label{S-tilde-manip}
\begin{split}
\vkS& = \frac{1}{|D|^2}(D^*\S_{+} + D \S_{-})\\
& = \frac{1}{|D|^2}(\S\Re[D] + \S_{*}\Im[D]),
\end{split}
\end{equation}

\noindent where we have used $D_- = D_+^*$ and written $D_+ = D$, and used the identity \ref{pm-hilbert-identity} between the first and second line.  Identity \ref{pm-hilbert-identity} also implies

\begin{align}
\Re[D] &= 1 - \pi c_s^2 G_*,\label{Re-D-eqn}\\
\Im[D] &= \pi c_s^2 G.\label{Im-D-eqn}
\end{align}

\noindent Note that an important property of $G_*$ is that it decays algebraically if $G(v)$ decays super-algebraically.  This is because the integral in Eqn.~\ref{hilbert-def}, which is dominated by $v \sim \vth$, may be expanded in powers of $k\vth/\Omega$ (in analogy to a multi-pole expansion to calculate the electric field far from a compact charge distribution).  Thus, Eqns.~\ref{S-tilde-manip}-\ref{Im-D-eqn} imply that if $\S(u)$ and $G(u)$ fall off faster than any power of $u$, then $\vkS(\Omega/k)$ falls off faster than any power of $\Omega/k$, as does the input power of any energy quantities formed by integrating $\mathcal{W}$ over $u$.  However, the quantity $\mathcal{W}$ is rather abstract -- what is its physical meaning?  It is also not clear precisely what Eqn.~\ref{Wcal-balance-eqn} implies for collisional dissipation.  Let us turn to a more conventional definition of energy.  Assuming $G$ has no zeros, Eqn.~\ref{plasma-eqn} with $S = 0$ conserves the quantity

\begin{equation}
W = \int vdv \frac{|f|^2}{\vth^2G} + \alpha |n|^2/2,
\end{equation}

\noindent where $\alpha = 2c_s^2/\vth^2$.  Now taking $G = 2v f_M/\vth^2$, this quantity is a reduced version\footnote{Eqn.~\ref{plasma-eqn} is reduced to one-dimension in velocity space.} of the free energy, which, for a gyro- or drift-kinetic plasma with a homogeneous Maxwellian background, is an exact collisionless global invariant when summed over ${\bf k}$.  The input of free energy $W$ also determines entropy production in the weakly collisional limit; see \ie \cite{krommes-hu, sugama96}.  It is obvious that if the input of $\mathcal{W}$ tends to zero, so does that of $W$.  However,  the input rate of $W$ can never be exactly zero, and so what is more important is how strongly it goes to zero.  It turns out that the input rate of $W$ goes to zero super-algebraically, for a general class of $S$.  In particular assuming $\S(v)  = S_0\varphi(v) f_M(v)$ where $\varphi = \sum \varphi_m v^m$, we can calculate the rate of free energy injection to be

\begin{equation}
\frac{dW}{dt} = \frac{2\pi}{k}|S_0|^2 \Re\left[H_+\left(\frac{\Omega}{k}\right)\right],\label{W-dot-eqn}
\end{equation}

\noindent where

\begin{equation}
\Re[H_+] = \Re\left[\frac{(\psi f_M)_+}{D_+} + \alpha \frac{f'_{M+}}{D_+}\Phi[\psi, f_M]\right],\label{H-plus-eqn}
\end{equation}

\noindent and $f'_M(v) = vf_M(v)$.  The details of this derivation are given in Appx.~\ref{nep-proof-appx}.  Because $\psi(v)$, given by Eqn.~\ref{psi-eqn}, is a purely real polynomial, identity \ref{pm-hilbert-identity} implies that this expression goes to zero super-algebraically in $k\vth/\Omega$.  To illustrative this, let us work out a simple case more explicitly, \ie two-moment forcing ($\varphi = 1 + c v$).  Using $f_{M+} = Z(\xi)/(2\pi i \vth)$, where $Z$ is the plasma dispersion function and $\xi = \Omega/(k\vth)$, we find

\begin{equation}
\frac{dW}{dt} = \frac{|S_0|^2}{k\vth}\Re\left[\frac{Z + (A + B \xi)[1 + \xi(1 + \xi Z)]}{i[1 + \alpha(1 + \xi Z)]}\right],\label{W-dot-2-moment}
\end{equation}

\noindent where $Z = Z(\xi)$, $A = 2 \vth \Re[c]$ and $B = \vth^2|c|^2(1 - \alpha + \alpha^2/(1+\alpha))$.  For large $\xi$, the Maxwellian case ($c = 0$) yields $dW/dt = |S_0|^2(1+\alpha)^2\sqrt{\pi}\exp(-\xi^2)/(k\vth)$, which can be compared with Fig.~\ref{fig:WdotScaling}.  

Note that on physical grounds the collisionless calculation performed in this section should agree with the weakly collisional limit (small but non-zero collision rate) of a plasma.  In this limit, energy input balances with dissipation in the steady state.  Therefore, we can interpret Eqn.~\ref{W-dot-eqn} as a measure of dissipation, and conclude that the plasma is behaving reversibly when this quantity is small.  In the following section, we will include dissipation explicitly and examine this scenario in more detail.

\input{jtp.tex}

\section{Summary and Discussion}

In Sec.~\ref{nep-sec} we proved a very general theorem about entropy production by a stable plasma Fourier mode.  In fact, it is a simple corollary of the completeness of linear eigenmodes, proved by \citet{case}, that the NEP theorem also applies to unstable wavenumbers when the system is projected into to the subspace of the stable continuum.  The continuum damping in these degrees of freedom can thus be separated from the energetics of the discrete modes, which exchange energy with the background in a fundamentally reversible way.  Consequently an individual Fourier mode of a plasma could avoid irreversible damping entirely via our mechanism.

In Sec.~\ref{hermite-sec} we approached the problem from the perspective of a weakly collisional plasma with a Hermite-space representation.  We found that the loss of damping is accompanied by a simultaneous super-algebraic decay of the Hermite spectrum and a loss of energy flow to small scales (large $m$).  Asymptotic analysis in the fast-forcing limit demonstrate how this occurs mathematically.  A practical outcome of this for physical systems with small but finite collisions is that, at modestly small values of $k\vth/\Omega$, the amount of dissipation occurring becomes small, and the small amount that does occur must happen at the largest scales (lowest Hermite index $m$); thus the ``cascade'' of energy to small scales can be effectively eliminated.

Collisionless plasma instabilities are sometimes described as ``inverse Landau damping.''  This is presumably due to the fact that a kind of ``phase-mixing'' occurs in the mechanics of an unstable plasma mode.  However, the terminology is unfortunate because it neglects the fundamental reasons that Landau modes are different from these discrete modes.  Discrete modes are smooth eigenmodes with velocity-space structure that is constant in time.  On the other hand, a Landau mode is a ``fake eigenmode'' (which is a superposition of stable eigenmodes), that secularly develops finer-and-finer scales in velocity space.  The NEP theorem suggests another, more physical feature that distinguishes these two classes of plasma modes: the continuum modes can avoid dissipation ``at all orders'' under detuning of the drive frequency, but discrete plasma eigenmodes will always gain or lose energy (without involving collisions) according to their eigenfrequencies.  This is remarkable because all roads lead to damping for the autonomous system ($S = 0$ in Eqn.~\ref{plasma-eqn}) considered by Landau -- that is, any reasonable initial condition will lead to Landau damping.  Considering the forced equation, however, we are led to the conclusion that Landau damping appears to be a delicate process, vulnerable to disruption.

\section*{Acknowledgements}

This work was supported by the UK Engineering and Physical Sciences Research Council through a Doctoral Training Grant award to J.T.P., with additional support from Award No KUK-C1-013-04 made by King Abdullah University of Science and Technology (KAUST).  G.G.P. acknowledges support from the Max-Planck/Princeton Research Center for Plasma Physics.


 \bibliographystyle{unsrtnat}
 \bibliography{nep-thm}

\appendix

\section{Kinetic \alf waves}\label{KAW-sec}

We now consider a slight variation of Eqn.~\ref{plasma-eqn}, also changing notation to conform with conventions.  Note that Appx.~\ref{KAW-sec} is independent of the rest of the paper, and so it should not be a significant problem if there are minor conflicts with notation that appears elsewhere.  The new equation is

\begin{equation}
\frac{\partial f}{\partial t} + i k v f + i k \left(n - \kappa \frac{v}{\vth} j\right) \alpha g = S,\label{plasma-kaw-eqn}
\end{equation}

\noindent where $g = v f_M$, the density is defined as before (Eqn.~\ref{n-eqn}), and the normalized parallel current is

\begin{equation}
j = \frac{1}{\vth}\int_{-\infty}^{\infty} vdv f.
\end{equation}

\noindent This is a single-kinetic-species model that is applicable to several physical regimes, depending on how the constants $\alpha$ and $\kappa$ are defined. For instance, it can be applied to describe a low-$\beta$ kinetic \alf wave (KAW) at scales similar to the ion Larmor radius, which is also similar to the electron inertia scale; see \citet{Zocco} for details.  In this case, $f$ represents the perturbed distribution for the electrons, while the ions respond adiabatically.  The equations could also describe electromagnetic kinetic ion dynamics near the ion inertia scale, with adiabatic electrons.  In any case, we are now considering a magnetized plasma with a time-independent and spatially-uniform guide field ${\bf B}_0 = B_0 \hat{\bf b}$ and a fluctuating parallel magnetic vector potential, \ie

\begin{equation}
j = \frac{1}{\kappa\alpha} \frac{q\delta A_{\parallel}}{T}\frac{\vth}{c}.
\end{equation}  

The analysis of Eqn.~\ref{plasma-kaw-eqn} is more tedious than than the analysis of Eqn.~\ref{plasma-eqn}, but it is fundamentally the same.  Indeed, the dimensionality of the problem is the same, \ie we are solving again for $f(k, v, t)$, and thus an analysis in the style of \citet{vk} should proceed analogously.  Let us do this analysis, since it appears to have not been done elsewhere, and also because it will demonstrate the procedure for deriving a transformation like that of Morrison.  Taking $S = 0$, we can write the eigenfunctions as

\begin{equation}
f^u(v) = -\alpha P\frac{g(v)}{v-u}\left(n^u - \frac{v}{\vth}\kappa j^u\right) + \lambda_u\delta(u-v).\label{pre-f-u-eqn}
\end{equation}

It is convenient to take the density moment of Eqn.~\ref{plasma-kaw-eqn} to obtain a relationship between $n^u$ and $j^u$:

\begin{equation}
j^u = -\frac{\gamma}{\kappa} \frac{u}{\vth} n^u,
\end{equation}

\noindent where $\gamma = 2\kappa/(\alpha\kappa - 2)$.  Then we take $n^u = 1$ as a choice of normalization.  This leads to

\begin{equation}
f^u(v) = -\alpha P\frac{g(v)}{v-u}\left(1 + \frac{uv}{\vth^2}\gamma\right) + \lambda_u\delta(u-v),\label{f-u-eqn}\\
\end{equation}

\noindent and

\begin{equation}
\lambda_u = 1 - \alpha\pi\left(1 + \frac{u^2}{\vth^2}\gamma\right)g_*,\label{lambda-eqn}
\end{equation}

\noindent where we have used identity \ref{hilbert-identity} to obtain Eqn.~\ref{lambda-eqn}.  Now we demonstrate that these eigenfunctions are complete, \ie that any function $f(v)$ can be written as

\begin{equation}
f(v) = \int_{-\infty}^{\infty} du C(u) f^u(v).\label{mt-kaw-inversion-formula}
\end{equation}

\noindent The task is to invert this expression for $C(u)$.  This will simultaneously (1) prove completeness of the eigenfunctions and (2) give us the transformation that we seek along with its inverse.  Substituting in Eqn.~\ref{f-u-eqn} we find

\begin{equation}
-\alpha\pi g\left[C_* + \frac{\gamma v}{\vth^2} C'_*\right] + \lambda_v C = f
\end{equation}

\noindent where we have notated $C'(v) = vC(v)$.  Using again identity \ref{hilbert-identity} and substituting in Eqn.~\ref{lambda-eqn} we obtain

\begin{equation}
\alpha g\frac{v}{\vth}\gamma\bar{C} - \alpha\pi\left(1 + \frac{v^2}{\vth^2}\gamma\right)\left [ gC_* + g_*C \right] + C = f,
\end{equation}

\noindent where we have introduced $\bar{C} = \vth^{-1}\int dv C(v)$.  Then we expand $f = f_+ + f_-$, $C = C_+ + C_-$, use the identity \ref{hibert-pm-identity}, and $v^2 g_*(v) = (v^2 g)_* + \vth^2/(2\pi)$.  This results in only terms which are positive- and negative-frequency functions, and so can be equated separately.  This yields the two equations

\begin{equation}
\pm2\pi i \alpha g'_\pm C_\pm + \left(1-\frac{\alpha\gamma}{2}\right)C_\pm + \bar{C}H_{\pm} = f_\pm,\label{pm-eqn}
\end{equation}

\noindent where we have defined $g' = (1 + \gamma v^2/\vth^2)g$ and $H = \alpha\gamma(1 + \gamma v^2/\vth^2)g/\vth$.  Solving Eqn.~\ref{pm-eqn} for $C_\pm$ we have

\begin{equation}
C_\pm = \frac{f_\pm - \bar{C}H_\pm}{D_\pm},\label{Cpm-soln-eqn}
\end{equation}

\noindent where we define

\begin{equation}
D_\pm = 1 -\alpha\gamma/2 \pm 2\pi i \alpha g'_\pm.
\end{equation}

What is left is to solve for $\bar{C}$, which can be done by integrating Eqn.~\ref{Cpm-soln-eqn} over $v$.  To express this cleanly, let us introduce the auxiliary transformation\footnote{The invertibility of the auxiliary transformation follows from the invertibility of Eqn.~\ref{mt-kaw-eqn}.  Furthermore, invertibility implies that it satisfies an equation analogous to Eqn.~\ref{mt-identity}, which is used between the first and second equality in Eqn.~\ref{cbar-eqn}.}

\begin{equation}
\breve{f} = \frac{f_+}{D_+} + \frac{f_-}{D_-}.
\end{equation}

\noindent Then, using $\int dv C(v) = 2\int dv C_\pm(v)$, Eqn.~\ref{Cpm-soln-eqn} leads to

\begin{equation}
\bar{C} = \frac{\frac{2}{\vth}\int dv \frac{f_\pm}{D_\pm}}{1 + \frac{2}{\vth}\int dv \frac{H_\pm}{D_\pm}} = \frac{\bar{\breve{f}}}{1 + \bar{\breve{H}}}.\label{cbar-eqn}
\end{equation}

Completeness of the eigenfunctions $f^u$ is thus proved (by Eqns.~\ref{Cpm-soln-eqn} and \ref{cbar-eqn}) if $C_+$ is a positive frequency function, which is ensured if the denominator in Eqn.~\ref{Cpm-soln-eqn} (\ie $D_+$) has no zeros in the upper half plane.  This is satisfied if and only if there are no unstable solutions to Eqn.~\ref{plasma-kaw-eqn}.  (This can be shown by Nyquist diagram, \ie by demonstrating both the dispersion relation and the denominator $D_+$ have the same image under mapping to the upper half plane.)  Finally, as promised, Eqn.~\ref{Cpm-soln-eqn} immediately provides the transformation via $\vkf(u) \equiv C(u) = C_+(u) + C_-(u)$, \ie

\begin{equation}
\vkf = \breve{f} - \breve{H}\frac{\bar{\breve{f}}}{1 + \bar{\breve{H}}},\label{mt-kaw-eqn}
\end{equation}

\noindent which can be inverted using Eqn.~\ref{mt-kaw-inversion-formula} by substituting $C(u) = \vkf(u)$.  This completes the derivation of the transformation for our system.  What is immediately implied is that Eqn.~\ref{plasma-kaw-eqn} is equivalent to Eqn.~\ref{plasma-eqn-trans}, where the transformation is the new one, Eqn.~\ref{mt-kaw-eqn}.  Thus, with the appropriate substitutions, Eqns.~\ref{f-large-t-eqn}-\ref{Im-D-eqn} will also apply, along with their physical implications.

\section{Identities}

A useful identity relating the Hilbert transform to the positive/negative frequency parts is

\begin{equation}
h_\pm = \frac{1}{2}(h \pm i h_*),\label{pm-hilbert-identity}
\end{equation}

\noindent implying also

\begin{equation}
h_* = -i(h_+ - h_-).\label{hibert-pm-identity}
\end{equation}

Integration of Eqn.~\ref{hilbert-def} by parts yields the identity

\begin{equation}
(vh)_* = v h_* - \frac{1}{\pi}\int_{-\infty}^{\infty} h(v') dv',\label{hilbert-identity}
\end{equation}

We will also need the following identity, which follows from the invertibility of the Morrison transform

\begin{equation}
(\vkf)_\pm = \frac{f_\pm}{D_\pm}.\label{mt-identity}
\end{equation}

\section{Derivation of rate of free energy input}\label{nep-proof-appx}

To compute the rate of free energy input at steady state, we start with the energy budget equation.  First we multiply Eqn.~\ref{plasma-eqn} by $f^*v/G = f^*\vth^2/(2f_M)$, integrate over velocity and take the real part.  Next we multiply Eqn.~\ref{plasma-eqn} by $\alpha n^*/2$, where $\alpha = 2c_s^2/\vth^2$, integrate over velocity, and take the real part.   The sum of the two equations yields

\begin{equation}
\frac{dW}{dt} = \Re\left[\e^{-i\Omega t}\left(\int\frac{f^* S(v)}{f_M}dv + \alpha n^*\int dv S(v)\right)\right].
\end{equation}

\noindent Now taking $S(v) = S_0 \varphi(v) f_M$, and using the long-time solution of $f$, Eqn.~\ref{f-large-t-eqn}, we obtain Eqn.~\ref{W-dot-eqn} with

\begin{equation}
H = \vks \left\{\alpha + \varphi^* + \alpha \pi\left(\left(\varphi^*f'\right)_* - \varphi^*f'_*\right)\right\}.\label{H-eqn}
\end{equation}

\noindent were $s = S(v)/S_0 = \varphi f_M$.  We will now manipulate Eqn.~\ref{H-eqn} in a series of steps, until we ultimately obtain Eqn.~\ref{H-plus-eqn}.  It is tedious but straightforward.  We will need a few identities along the way.  First, for any $h(v)$ and $\varphi = \sum_{m=0}^{N} \varphi_m v^m$ we have

\begin{equation}
(\varphi h)_* = \varphi h_* - \frac{1}{\pi}\Phi[\varphi, v h],\label{identity-1-eqn}
\end{equation}

\noindent where we define the polynomial

\begin{equation}
\Phi[\varphi, h](u) = \sum_{m=0}^{N-1} T_{m+1}[\varphi]\frac{1}{u^{m+1}}\int dv h(v) v^m,
\end{equation}

\noindent and the truncated polynomial

\begin{equation}
T_m[\varphi](u) = \sum_{n=m}^N \varphi_n u^n.
\end{equation}

Now using Eqn.~\ref{identity-1-eqn} we may rewrite Eqn.~\ref{H-eqn} as

\begin{equation}
H = \vks\left\{\varphi^* + \alpha \left(1 - \Phi[\varphi^*, f_M']\right)\right\},\label{H-eqn-2}
\end{equation}

\noindent where $f_M'(v) = v f_M$.  It can be shown that

\begin{equation}
1 - \Phi[\varphi^*, f_M'] = \varphi^* - u \Phi^*,\label{Phi-identity-eqn}
\end{equation}

\noindent where we have introduced the abbreviation $\Phi[\varphi, f_M] = \Phi$.  Eqn.~\ref{H-eqn-2} now becomes

\begin{equation}
H = \vks\left\{(1+\alpha)\varphi^* - \alpha u \Phi^* \right\}.\label{H-eqn-3}
\end{equation}

\noindent Now from Eqn.~\ref{identity-1-eqn} (and Eqn.~\ref{pm-hilbert-identity}) we find 

\begin{equation}
\vks = \widetilde{f_M\varphi} = \vkf_M\varphi - \frac{i}{2\pi} Q P,\label{Q-intro-eqn}
\end{equation}

\noindent where

\begin{align}
Q& = \frac{1}{D_+} - \frac{1}{D_-},\label{Q-eqn-1}\\
& = -2\pi i \alpha \vkf_M',\label{Q-eqn-2}\\
& = -2\pi i \alpha u \vkf_M/(1+\alpha).\label{Q-eqn-3}
\end{align}

\noindent Using Eqns.~\ref{Q-intro-eqn} and ~\ref{Q-eqn-3} then leads to the form

\begin{equation}
H = \psi(u) \vkf_M,\label{H-eqn-4}
\end{equation}

\noindent where

\begin{equation}
\psi = (1+\alpha)|\varphi|^2 + u^2\frac{\alpha^2}{1 + \alpha}|\Phi|^2 - \alpha u(\varphi\Phi^* + \varphi^*\Phi),\label{psi-eqn}
\end{equation}

\noindent is a purely real polynomial.  Now, using Eqn.~\ref{identity-1-eqn}, we can write

\begin{equation}
H = \widetilde{\psi f_M} + \frac{i}{2\pi}Q\Phi[\psi, f_M].\label{H-eqn-5}
\end{equation}

\noindent Finally, we take the positive frequency part of this equation, apply Eqn.~\ref{mt-identity} and Eqn.~\ref{identity-1-eqn} once more, and find

\begin{equation}
H_+ = \frac{(\psi f_M)_+}{D_+} + \frac{i}{2\pi}Q_+\Phi[\psi, f_M] - \left(\frac{i}{2\pi}\right)^2\Phi[\Phi[\psi,f_M], Q].\label{H-eqn-6}
\end{equation}

\noindent Because $Q$ is purely imaginary, the final term can be neglected in computing $\Re[H_+]$.  Using Eqn.~\ref{Q-eqn-2} we then obtain Eqn.~\ref{H-plus-eqn}.  

\end{document}

%% file: jtp.tex
\section{Formulation in Hermite Space}\label{hermite-sec}

To facilitate numerical solution and to provide a different mathematical viewpoint, let us reformulate the problem as weakly collisional dynamics in Hermite space.  
We add a dissipation operator to the right hand side of Eqn.~\ref{plasma-eqn} and represent our distribution function with a Hermite series

\begin{align}
  f(v,t) &= \sum_{m=0}^{\infty} a_m(t)  \frac{H_m(\hat{v})}{\sqrt{2^m m!}} f_M ,
\end{align}

\noindent where $\hat{v}=v/\vth$ and the Hermite polynomials are

\begin{align}
  H_m(\hat{v}) &= (-1)^m e^{\hat{v}^2}\frac{d^m}{d \hat{v}^m}\left( e^{-\hat{v}^2}\right).
\end{align}

This is a popular approach for solving kinetic systems \citep{Grad49, GradNote, Grant, Grant67PhysFluids}, which gives an explicit definition of scale via the Hermite index $m$.  
The kinetic equation is now
\begin{align}
  \begin{split}
  \frac{\partial a_m}{\partial t} 
  &+ ik\vth\left( \sqrt{\frac{m+1}{2}}a_{m+1} + \sqrt{\frac{m}{2}}a_{m-1} \right)
  \\
   &
   \hspace{1cm}
   + ikc_s^2 G_m a_0 
  = S_m e^{-i\Omega t} - \nu m^n a_m 
  \end{split}
  \label{eq:HermiteKineticEquation}
\end{align}

\noindent where we have expanded $S$ and $G$ in Hermite moments (where $G_m=\sqrt{2}\delta_{m1}/\vth$), and included an $n$-iterated L\'enard--Bernstein collision operator \citep{Joyce71, hatch-2013}.
Linear phase mixing now takes the form of nearest-neighbor coupling in $m$-space, and we may track the flow of energy to ``fine scales'' (high-$m$) by directly examining these couplings and inspecting the steady-state $m$-spectrum.

\subsection{Entropy production}

The free energy balance can now be written

\begin{align}
  \frac{dW}{dt}
  &= \dot{W}_S + \dot{W}_C,
  \\
  W &= \frac{\alpha}{2}|a_0|^2 + \frac{1}{2}\sum_{m=0}^{\infty}|a_m|^2
  \\
  \dot{W}_S
  &= \Re\left( \alpha a_0^* S_0 e^{-i\Omega t} + \sum_{m=0}^{\infty} a_m^* S_me^{-i\Omega t} \right),
  \label{eq:FreeEnergyInjection}
  \\
  \dot{W}_C 
  &= - \nu \sum_{m=0}^{\infty} m^n|a_m|^2 .
  \label{eq:FreeEnergySink}
\end{align}

Let us consider what happens to our system under frequency detuning $\omegaNL > k\vth$.

\subsection{Numerical comparison and asymptotic solution}

The verify Eqn.~\ref{W-dot-eqn} and the NEP theorem, let us consider numerical solutions of Eqn.~\ref{eq:HermiteKineticEquation}, truncating the Hermite expansion for $f$ after $N$ terms.  We consider the numerical solutions to be resolved when two conditions are met: (1) $N$ is sufficiently large that the high-$m$ dissipation is completely resolved, and further increase in $N$ does not change the the rate of consumption of free energy and (2) dissipation $\nu$ is small enough such that low-$m$ dissipation is negligible compared to high-$m$ dissipation.  The second condition assures that the solution will agree with the collisionless result.

\begin{figure}
\resizebox{0.5\textwidth}{!}{%
  \includegraphics{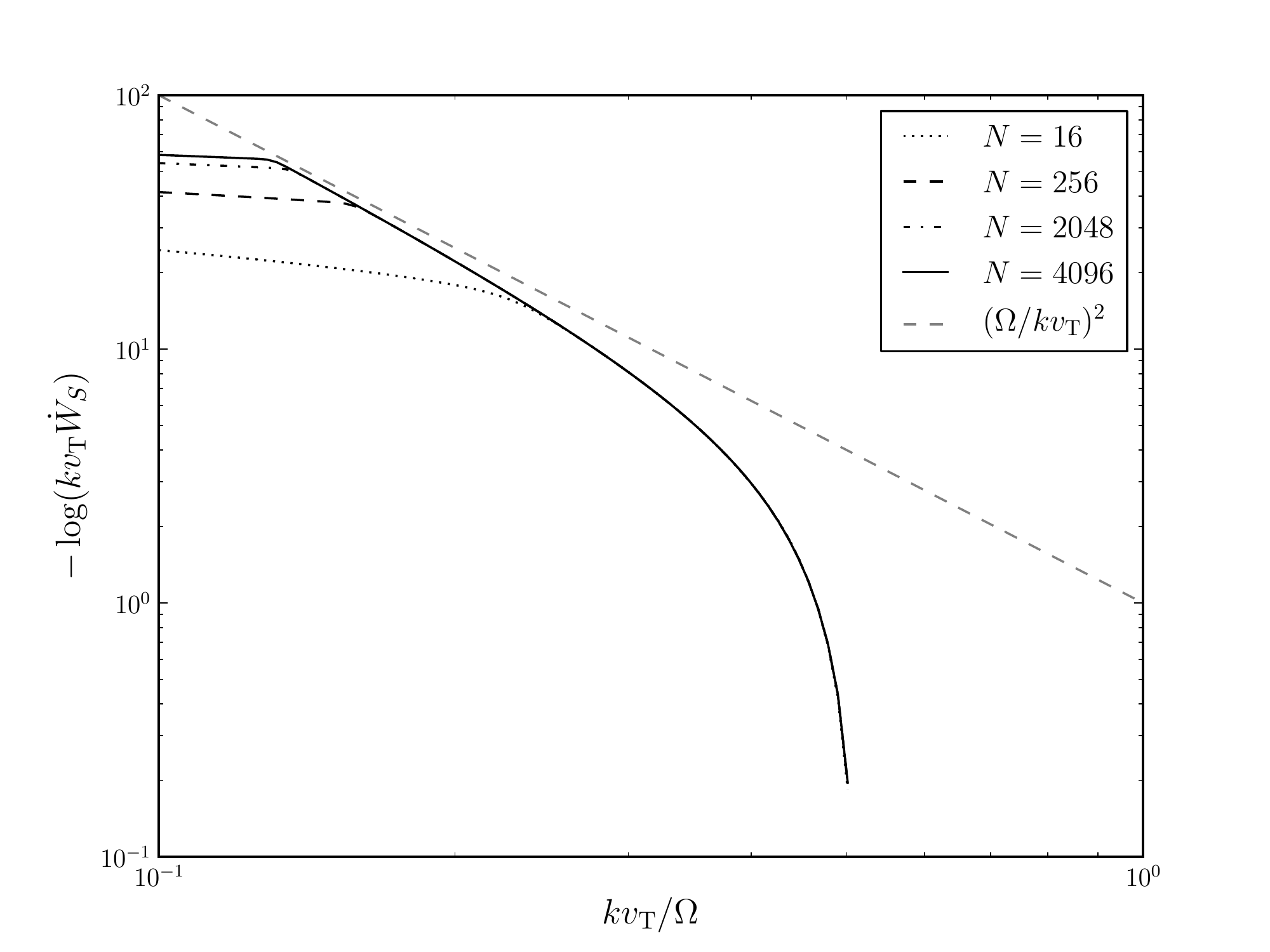}
}
\caption{Free energy injection against $k\vth/\Omega$ for $\alpha = 1$ and various resolutions $N$.
}
\label{fig:WdotScaling}
\end{figure}

In Fig.~\ref{fig:WdotScaling} we plot the long-time result $-\log(|k\vth\dot{W}_S|)$ against $k\vth/\Omega$ for the Maxwellian driving case $S_m=\delta_{m0}$.  This has gradient $-2$ in the limit $k\vth/\Omega\to0$, indicating that $\dot{W}_S\sim \exp(-(\Omega/k\vth)^2)/k\vth$, in accordance with Eqn.~\ref{W-dot-2-moment}.  In general we may take any set of $S_m$, but the Maxwellian driving case is representative.

\begin{figure}
\resizebox{0.5\textwidth}{!}{%
  \includegraphics{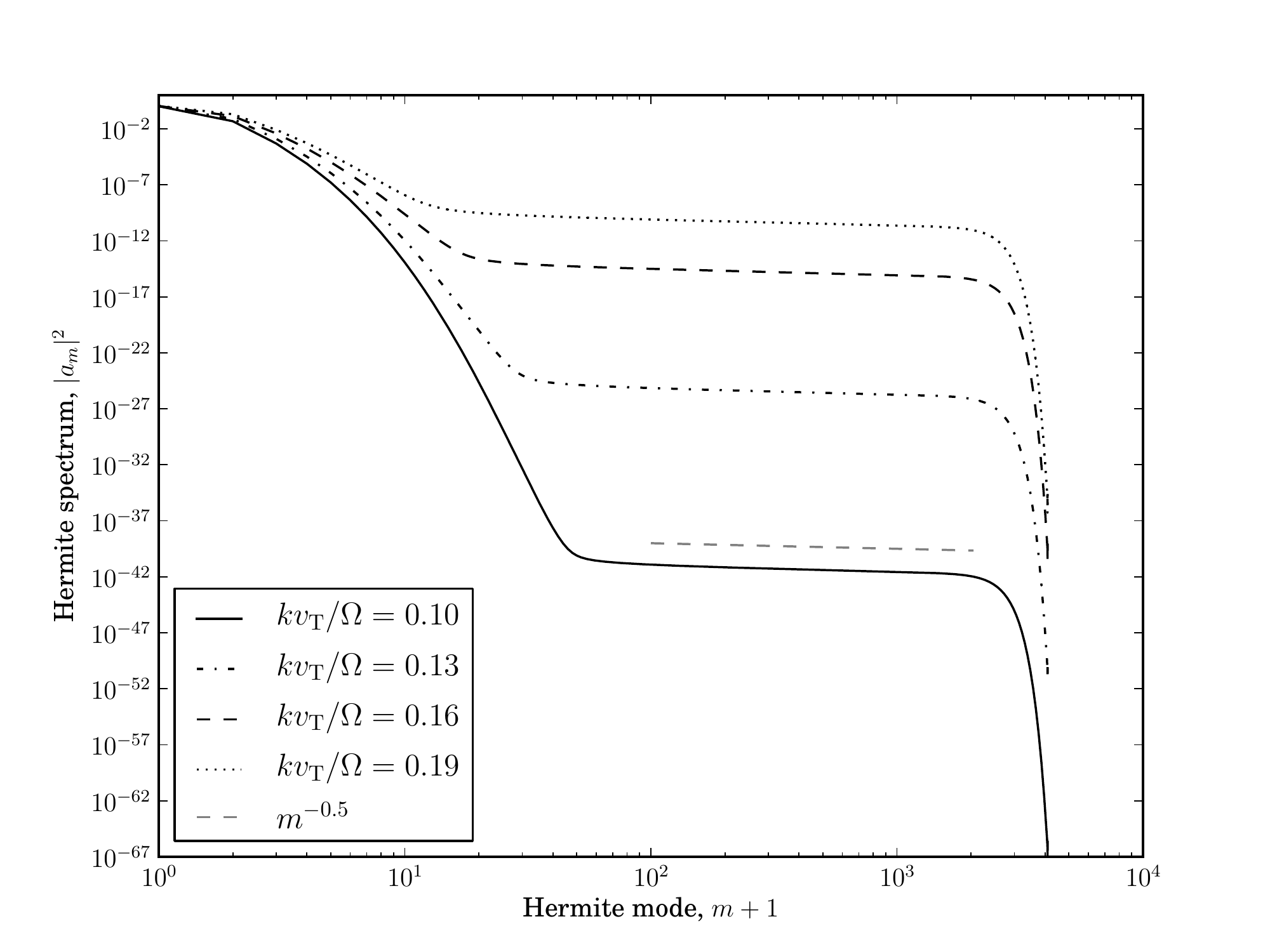}
}
\caption{
Hermite spectra for $N=4096$ and $\alpha = 1$ at different $k\vth/\Omega$.
}
\label{fig:HermiteSpectrum}
\end{figure}

Each resolution $N$ in Fig.~\ref{fig:WdotScaling} has a critical $k\vth/\Omega$ below which the scaling does not hold, because the collisionless behavior is not resolved and large-scale (low-$m$) dissipation becomes important.  The simple explanation for this is that dissipation $W_C$ (Eqn.~\ref{eq:FreeEnergySink}) depends on the Hermite spectrum, which is insensitive to $k\vth/\Omega$ at low-$m$, but decays very sharply at high-$m$ as $k\vth/\Omega \to 0$ (see Fig.~\ref{fig:HermiteSpectrum}). Therefore, although it is always possible in principle to resolve the collisionless behavior, it becomes very costly at small $k\vth/\Omega$ because $\nu$ and $N$ must be extraordinarily small and large respectively, which is prohibitive for our numerical scheme.  The ``corner'' in Fig.~\ref{fig:WdotScaling} thus marks the point of transition where the low-$m$ and high-$m$ contributions to $\dot{W}_C$ balance.

Now let us consider the Hermite spectra.  These are shown in Fig.~\ref{fig:HermiteSpectrum} for various $k$ in the Maxwellian driving case.
There are three distinct behaviors corresponding to low, medium and high $m$. 
The medium and high $m$ case was described by \citet{Zocco}.
For $m\gg1$ one may neglect driving and the Boltzmann response, and derive an energy equation from \ref{eq:HermiteKineticEquation} where the streaming term is approximated with an $m$-derivative.
Solving this gives the spectrum for large $m$,
\begin{align}
  |a_m|^2 \propto \frac{1}{\sqrt{m}}\exp\left( -\frac{\nu}{k\vth}\frac{\sqrt{2}m^{n+1/2}}{2n+1}\right)
  \label{eq:LargeMSpectrum}
\end{align}

\noindent which is in excellent agreement with the spectra in Fig.~\ref{fig:HermiteSpectrum}.
For medium $m$ the spectrum behaves like $1/\sqrt{m}$, while for large $m$ the exponential collision term dominates.  Thus it is unavoidable that the flow of energy ultimately assumes a conventional form of the autonomous (Landau) solution at sufficiently large $m$.  However, the amount of dissipation depends on the overall amplitude of this Landau tail, which is determined by the small-$m$ behavior of the spectrum.

To gain insight into behavior at small $m$, we can solve the kinetic equation \ref{eq:HermiteKineticEquation} neglecting collisions ($\nu=0$) in the long-time limit, \ie taking $a_m(t)=\hat{a}_me^{-i\Omega t}$. 
We assume the source only drives the first $\bar{m}$ 
moments, such that $S_m=0$ for all $m>\bar{m}$. 
Eqn. \ref{eq:HermiteKineticEquation} becomes

\begin{align}
  -i\Omega \hat{a}_m + i k \vth \L \hat{a}_m = S_m,\label{hermite-ss-kinetic-eqn}
\end{align}

\noindent where $\L$ is the operator

\begin{align}
  \L \hat{a}_m = 
  \sqrt{\frac{m+1}{2}} \hat{a}_{m+1}  
  + \sqrt{\frac{m}{2}}(1+\alpha\delta_{m1}) \hat{a}_{m-1}.
  \label{eq:StreamingD}
\end{align}


Let us consider the limit $k\vth\sqrt{m}/\Omega\ll1$, for which we can make analytical progress.
We solve Eqn.~\ref{hermite-ss-kinetic-eqn} treating $k\vth/\Omega$ as a small parameter.  
Setting 

\begin{equation}
\hat{a}_m=\sum_{n=0}^{\infty}\hat{a}_m^{(n)}(k\vth/\Omega)^n,
\end{equation}

\noindent we find that streaming vanishes at leading order and the equation is local in $m$ with the solution 

\begin{align}
  \hat{a}_m^{(0)} = 
  \frac{iS_m}{\Omega}.
\end{align}
From Eqn.~\ref{hermite-ss-kinetic-eqn}, higher orders are found iteratively with
\begin{align}
  \hat{a}_m^{(n)} = 
  \L \hat{a}_m^{(n-1)},
  \hspace{1cm} 
 & n \geq 1,
  \label{eq:OrderByOrderSoln}
\end{align}

\noindent Thus we immediately find at zeroth order that
$\hat{a}_m = iS_m/\Omega$ 
for the forced components $m \leq \bar{m}$, 
while the unforced components, $m > \bar{m}$, are zero.
Moreover as coupling is nearest-neighbor, 
every moment with $m>\bar{m}$ 
is an order higher in $k\vth/\Omega$ than the moment below it, \ie

\begin{equation}
\begin{split}
&\hat{a}_{\bar{m}+1} = \frac{k\vth}{\sqrt{2}\Omega}\sqrt{\bar{m}+1}(1 + \alpha\delta_{\bar{m}0})\hat{a}_{\bar{m}},\\
&\hat{a}_{\bar{m}+2} = \frac{k^2\vth^2}{2\Omega^2}\sqrt{(\bar{m}+2)(\bar{m}+1)}(1 + \alpha\delta_{\bar{m}0})\hat{a}_{\bar{m}},
\end{split}
\end{equation}

\noindent and so forth, implying $\hat{a}_m \sim {\mathcal O}(\hat{a}_{\bar{m}}(k\vth/\Omega)^{m-\bar{m}}\sqrt{m!/\bar{m}!})$.  Thus, $\hat{a}_m$ falls off rapidly 
with $m$
until $k\vth\sqrt{m}/\Omega \sim O(1)$ and streaming can no longer be neglected at leading order.
The point of transition to the $m^{-1/2}$ spectrum in Fig.~\ref{fig:HermiteSpectrum} agrees well with $k\vth\sqrt{m}/\Omega\sim O(1)$.

This solution also illustrates that free energy injection $\dot{W}_S$ vanishes at every order in $k\vth/\Omega$.
Let us first show this for $\alpha=0$, and then generalize the solution.  The $n$th-order contribution to $\dot{W}_S$ is
\begin{align}
  \dot{W}_S^{(n)} = \Re\left( \sum_{m=0}^{\infty} \hat{a}_m^{(n)*}S_m\right).
\end{align}
The $n=0$ case trivially vanishes, while for $n\geq1$
\begin{align}
  \dot{W}_S^{(n)} = \Re\left( \frac{i}{\Omega}\sum_{m=0}^{\infty} (\L^n S_m^*)S_m\right).
\end{align}
By shifting the indices in the sum we may show that
\begin{align}
  \sum_{m=0}^{\infty} (\L^{n} S_m^*)S_m
 =
 \sum_{m=0}^{\infty} (\L^{n-1} S_m^*)\L S_m,
  \label{eq:Shift}
\end{align}
so that iterating
\begin{align}
  \dot{W}_S^{(n)} = \Re\left( \frac{i}{\Omega}\sum_{m=0}^{\infty} S_m^* \L^nS_m\right),
\end{align}
and $\dot{W}_S^{(n)}$ therefore vanishes as the sum is equal to its complex conjugate.

This proof relies on Eqn.~\ref{eq:Shift} which follows from the symmetry that the two terms in $S_m\L S_m^*$ are complex conjugates but with a shift in $m$. 
This symmetry is broken when $\alpha\neq0$, but may be restored with the substitutions
\begin{align}
  b_m =
  \begin{cases}
    \hat{a}_0\sqrt{1+\alpha} , \hspace{1cm} & m = 0,
    \\
    \hat{a}_m, \hspace{1cm} & \textrm{else},
  \end{cases}
  \\
  S'_m =
  \begin{cases}
    S_0\sqrt{1+\alpha} , \hspace{1cm} & m = 0,
    \\
    S_m, \hspace{1cm} & \textrm{else}.
  \end{cases}
\end{align}
The kinetic equation \ref{hermite-ss-kinetic-eqn} becomes
\begin{align}
  -i\Omega b_m + i k \vth \L'b_m = S'_m,
\end{align}

\noindent where $\L'$ is the symmetric operator

\begin{align}
  \L'b_m = \sqrt{\frac{m+1}{2}}(1+\beta\delta_{m0})b_{m+1}
  +  \sqrt{\frac{m}{2}}(1+\beta\delta_{m1})b_{m-1},
\end{align}
(with $\beta = -1 + \sqrt{1+\alpha}$) which also satisfies Eqn.~\ref{eq:Shift}.  
Furthermore, free energy injection, Eqn.~\ref{eq:FreeEnergyInjection}, becomes simply
\begin{align}
  \dot{W}_S = \Re\left(\sum_{m=0}^{\infty} b_m^*S'_m \right).
\end{align}
Thus by the same argument as before, the free energy injection in the general $\alpha\neq0$ case vanishes at every order in $k\vth/\Omega$.

